\documentclass[12pt]{article}
\pdfoutput=1

\usepackage{graphicx}
\usepackage{verbatim}

\usepackage{times}

\usepackage{amsmath}
\usepackage{amsfonts}
\usepackage{mathtools}

\title{The Emergence of Innovation Complexity at Different Geographical and Technological Scales}

\author
{Emanuele Pugliese$^{1}$, Lorenzo Napolitano$^{1,2}$, Matteo Chinazzi$^{3}$, \\ 
Guido L. Chiarotti$^{1,4}$\\ 
\\
\normalsize{$^{1}$Institute for Complex Systems, Consiglio Nazionale delle 
Ricerche - Rome, Italy.}\\
\normalsize{$^{2}$Institute of Economics, Scuola Superiore Sant'Anna - Pisa, 
Italy.}\\
\normalsize{$^{3}$MOBS Lab, Network Science Institute, Northeastern University - Boston, MA, USA.}\\
\vspace{-5pt}\normalsize{$^{4}$Department of Environmental, Land and Infrastructure Engineering,}\\
\vspace{17pt}\normalsize{Politecnico di Torino - Torino, Italy.}
\\
}

\date{}

\begin{document} 



\maketitle

\begin{abstract}

We define a novel quantitative strategy inspired by the ecological notion of \emph{nestedness} to single out the scale at which innovation complexity emerges from the aggregation of specialized building blocks.
Our analysis not only suggests that the 
innovation space 
can be interpreted as a natural system in which advantageous capabilities are selected by evolutionary pressure, but also that the emerging structure of capabilities is not independent of the scale of observation at which they are observed.
Expanding on this insight allows us to understand whether the capabilities characterizing the 
innovation space 
at a given scale are compatible with a complex evolutionary dynamics or, rather, a set of essentially independent activities allowing to reduce the system \emph{at that scale} to a set of disjoint non interacting sub-systems.
This yields a measure of the \emph{innovation complexity} of the system, \emph{i.e.} of the degree of interdependence between the sets of capabilities underlying the system's building blocks.

\end{abstract}
\pagebreak

\section{Introduction}

We define a novel quantitative strategy to single out the scale at which irreducible diversification -- \emph{i.e. complexity} -- emerges from the aggregation of specialized building blocks. 
We then apply this newly defined methodology to study the 
innovation space
as described by patented inventions and show the existence of a non-trivial interaction between the geographical and technological scales at which  \emph{innovation complexity} emerges. 
To capture the emergence of innovation complexity, we leverage on a property shared by many ecological and socio-economic systems: \emph{nestedness}.

In many ecosystems, populations of pollinator insects and plants coexist,  some of which are specialized while others are more generalist.
A commonly observed characteristic of these ecosystems is that specialist insects, which pollinate few species of plants, tend to prefer mostly generalist plants, which in turn are targeted by many different pollinator species, both specialist and not \cite{bascompte2003nested}.
The resulting general lack of specialist-specialist interactions is called nestedness and it is addressed by several ecological studies about the distribution of communities of species in geographically accessible sites \cite{patterson1986nested,wright1992meaning,cutler1991nested,cutler1994nested,wright1997comparative} and the structural organization of species interaction networks \cite{bascompte2003nested,dupont2003structure,ollerton2007finding,ollerton2003pollination,guimaraes2006asymmetries,guimaraes2007nested,lewinsohn2006structure,burns2007network}.

Nestedness can be rephrased in terms of the way in which traits are selected by evolutionary pressure. For instance, if we observed  specialist insects interacting mostly with specialist plants, we could infer that characteristics (i.e. capabilities) which benefit foraging from -- or being pollinated by -- some species tend to hinder interaction with others; this would translate to a plant-pollinator system consisting of non-interacting sub-systems. Nestedness implies that capabilities are not mutually exclusive, but rather that they imply each other through cumulative causation. Consequently, a nested plant-pollinator system will consist of non-separable sub-systems characterized by non-independent capability sets and emerging complexity.

Interestingly, the complexity in the structure of capabilities that emerges from the observation of an ecological niche at a scale of observation might not emerge at different scales.
For example, zooming into the social structure of honeybee communities highlights a strong division of labor involving specialization and non-overlapping capabilities between the fertile but dependent queen, the female workers, and the stingless male drones.
Indeed, the nested capabilities making honeybees generalist pollinators emerge only via aggregation from the compartmentalized capabilities behind their social cast structure and in the framework of the interaction between their social structures and the rest of the ecosystem.

Nestedness has been shown to characterize also social systems \cite{hidalgo2007product, hausmann2011network}, allowing to rank their agents on the basis of their fitness \cite{tacchella2012new}.
Similarly, in the innovation space
the presence of nestedness in the network of relations linking geographical areas hosting inventors to the technologies embedded in their patented inventions is a sign that more complex activities imply the capability to successfully perform simpler ones.
This implies the presence of a complex and irreducible system of interactions, in contrast with the intuition of mainstream models of economic specialization.
In a system characterized by clustered dependencies between technologies (i.e. low nestedness) innovators could afford to specialize in a small set of arbitrary patents.
Conversely, in a system governed by complex interactions between aggregates, the nested nature of the set of capabilities would force the fittest agents to treat simpler activities as stepping stones to master more complex innovations.
In this paper we show that this is precisely the case when the system is observed at the appropriate scale.
This way, we aim to inform the rich debate economic literature concerning the role of knowledge spillovers in fostering innovation \cite{boschma2014relatedness, boschma2014scientific, balland2015proximity, balland2017geography} by pointing out the role that the geographical scale of the analysis plays in driving empirical results and eventually contribute to defining a solid empirical basis \cite{frenken2017economic} for future theoretical as well as policy-making efforts.

There is a rich theoretical literature describing the 
innovation space as a complex system in which the intricate interconnectedness between fields of knowledge rises both directly -- from the recombination of concepts from different fields and the influence that innovations have on collective behavior and thinking patterns \cite{strumsky2012using, tria2014dynamics} -- and indirectly -- through the multiplicative effect that some innovations (\emph{e.g.} in computing, instrument building, information communication technology) have on the potential to innovate in unrelated fields \cite{bresnahan1995general, napolitano2018technology}.
Our methodology adds to this insight by measuring the nestedness of the innovation system at multiple resolutions and showing that the scale of observation not only matters, but that there is a non-trivial interaction between geographical and technological scale. Such interaction determines the emergence of a well-defined frontier separating the scale of observation characterized by diversification of the fittest (high nestedness) from the one displaying evolutionary pressure towards specialization (low nestedness).
To our best knowledge, this is the first time that the geographical scale at which complexity emerges is identified in a quantitative manner.

\section{Representing the Innovation Space}

Patent applications, being one of the main sources of codified information concerning inventions \cite{griliches1990patent}, are an ideal source of data for our analysis because the
associated metadata provides us with a standardized classification system -- the International Patent Classification (IPC) scheme\footnote{http://www.wipo.int/classifications/ipc/en/} -- that allows to associate the claims of innovativeness of each invention to one or more \emph{technological fields}. 
Furthermore, we are able to link patent inventors or assignees to their physical residence (or seat) at the time the patent application was filed
and therefore to assign codes to specific geographical regions.
This allows us to represent 
the innovation space
as a bipartite graph connecting two types of nodes: geographical areas and technological fields
.

\begin{figure}
 \centering
 \includegraphics[width=0.8\textwidth]{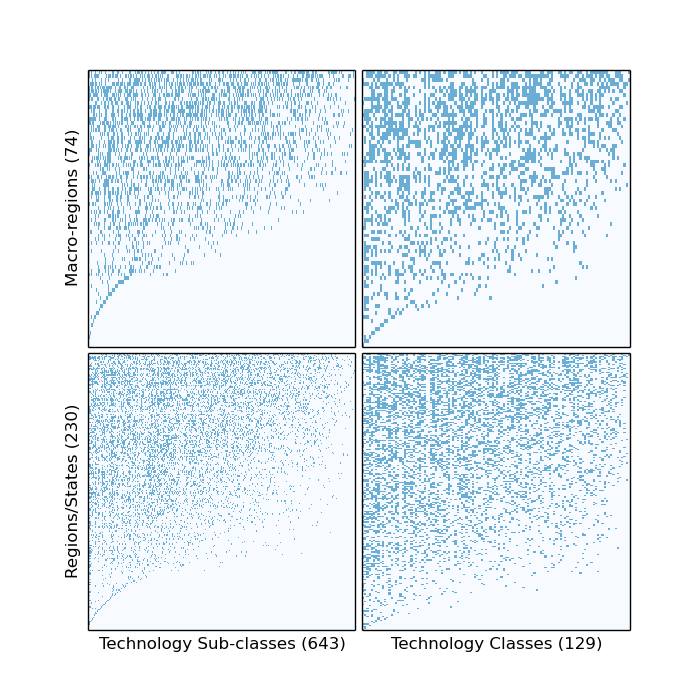}
 \caption{
\textbf{Innovation Maps at different scale combinations.}
This figure, reports some \emph{Innovation Maps} ($IMs$) at different levels of resolution (zooms), both along the geographical and technological dimensions. 
The relation matrices are built by joining information about the technological content of patents (IPC Codes) and the seat (or residence) of patentees
.
The coarsest $IM$ reported in this figure (top-right) consists of 74 rows and 129 columns, while the highest resolution $IM$ (bottom-left) represents 230 regions and 643 technological classes. 
The matrices have been reordered to emphasize nestedness by means of the algorithm proposed by \cite{tacchella2012new}. 
Note that the nonzero entries (colored dots) are concentrated in the top-left corner at all scales, though to different extents, suggesting the presence of widespread nestedness throughout the system.}\label{fig:4scales}
\end{figure}

The geographical and technological dimensions thus constructed are inherently hierarchical and, hence, suitable for the exploration of the data at the variety of scale combinations required for our investigation.
For all available scale combinations, we use IPC codes to decompose patents into sets of technological fields and assign patentees to their geographical regions. 
An equal share of each invention is attributed to all the involved location-technology pairs. 
Each pair maps to a cell of the so-called \emph{Innovation Map} ($IM$) which represents the 
innovation space 
at a given scale combination the same way in which geographic maps represent the Earth surface at different levels of resolution. 
Innovation maps can also be interpreted as incidence matrices of a bipartite network in which rows correspond to geograhical locations, columns correspond to technology fields, and entries are nonzero whenever the corresponding location-technology combination displays a revealed competitive advantage
\footnote{The results are robust to different binarization strategies.} 
and is thus well-equipped to innovate in a specific technological field. 
Figure \ref{fig:4scales} depicts some examples of IMs computed at specific geographical and technological resolutions. 
Given these representations of the 
innovation space
, we measure their nestedness \cite{rodriguez2006new} to uncover the footprint of \emph{innovation complexity}.

\section{Shuffling Capabilities: a Null Model}\label{sec:srl}

In order to test the statistical significance of the characteristic nestedness of a specific scale of observation, we need a suitable null hypothesis accounting for the fact that some nestedness carries over to any random rearrangement of the data contained in the IMs from lower resolutions. As a result, the excess nestedness we measure at any scale is a result of enhancing the resolution of data representation.
To this aim, we contrast the nestedness of a high resolution empirical innovation map, $IM^{highRes}$, with the expected nestedness given the distribution of technological capabilities observed in its lower resolution counterpart $IM^{lowRes}$.
In particular, we generate a null distribution of nestedness values from a set of synthetic IMs of the same shape as $IM^{highRes}$ which preserve the distribution of technological competences within $IM^{lowRes}$.
If the nestedness of $IM^{highRes}$ is not significantly different to that implied by the null distribution, we deduce that zooming-in brings out no further nestedness with respect to $IM^{lowRes}$ and hence highlights no further structural property of the system.
On the other hand, finding a significantly higher or lower nestedness than the null distribution suggests that $IM^{highRes}$ is either significantly more or significantly less nested than implied by $IM^{lowRes}$ and hence that changing scale yields information about the system.

Put into the context of the 
innovation space, our null hypothesis implies that innovative capabilities relative to a specific technological field that are observed in a given geographical area (\emph{e.g.} a nation) are randomly shuffled between the child sub-regions of said area (e.g regions or states).
The resulting null model -- which we call Reshuffled Capabilities model (ReCap) -- allows to define a distribution of nestedness values that are compatible with the nestedness measured at a coarser geograhical resolution. 
Assessing the significance of the nestedness of $IM^{highRes}$ with respect to the null distribution amounts to washing out the nestedness of $IM^{lowRes}$ from $IM^{highRes}$ thus allowing to test the alternative hypothesis that observing the system at a higher resolution uncovers unexpectedly high (or low) nestedness in the geographical structure of technological capabilities (which we interpret as a signal of irreducible diversification) against the null hypothesis that the nestedness observed in $IM^{highRes}$ is nothing more than the residual of that already measured in the more aggregated matrix.

The ReCap fixes the ubiquities of the technological codes (\emph{i.e.} the column sums) within $IM^{highRes}$ while letting the locations (i.e the row sums) vary freely within the blocks of rows corresponding to the rows of $IM^{lowRes}$.
In other words, building a null matrix from $IM^{highRes}$ amounts to splitting it into its constituent sub-matrices, shuffling each one according to the null model, and then recomposing the mosaic. 
The nestedness temperature \cite{atmar1993measure,rodriguez2006new} of the null matrices is measured 
and used to construct the null distribution, against which the nestedness temperature of $IM^{highRes}$ is compared to assess significance.

\begin{figure}
 \centering
 \includegraphics[height=0.83\textheight]{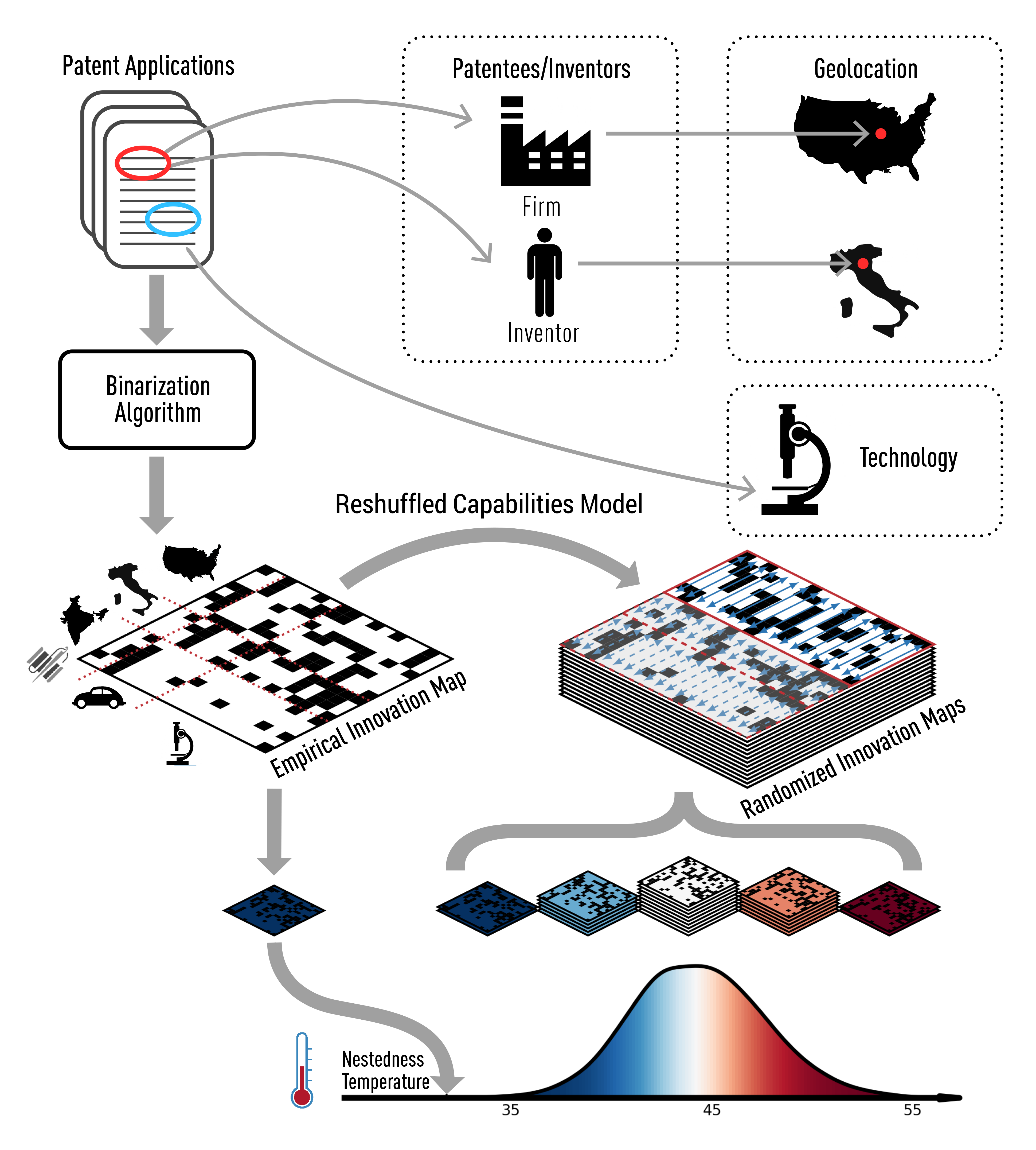}
 \caption{\textbf{From Patents to Nestedness Temperature}
This is a schematic representation of the overall procedure starting from collecting the geographical and technological information from patent applications to applying the Reshuffled Capabilities (ReCap) null model to measure the significance of observed nestedness. 
In this example, the empirical matrix is blue because its measured nestedness temperature is significantly lower (around $3\sigma$) than the average null matrix, implying that it is significantly more nested. }
\label{fig:explanation}
\end{figure}

Figure \ref{fig:explanation} contains a schematic representation of the overall procedure starting from collecting the geographical and technological information from patent applications to applying ReCap to measure the significance of observed nestedness.

\section{Results}

Our analysis shows that, for a given technological disaggregation, zooming to a finer geographical scale produces a structure that is less and less significantly nested with respect to a random reallocation of technological competitiveness among the geographical subunits of the system represented at the coarser scale. 
Eventually, zooming-in along the geograhical dimension of the system yields a matrix that is consistent with a process fostering specialization over diversification. 
This implies that the less common and more complex technology codes are predominantly targeted by firms localized in specialized areas when such areas are sufficiently narrow. 
For example, when we consider technologies at the IPC sub-class scale, we observe two regimes.
When the geographical scale is set to identify nations as units of analysis, the individual geographical areas display a more-than-random level of diversification (i.e. the 
innovation space
is nested). 
Instead, when we \emph{zoom-in} at the regional level of disaggregation, specialization emerges and we find that rare technologies are pursued in specialized industrial districts (i.e. the 
innovation space
is anti-nested).

\begin{figure}
 \centering
 \includegraphics[width=\textwidth]{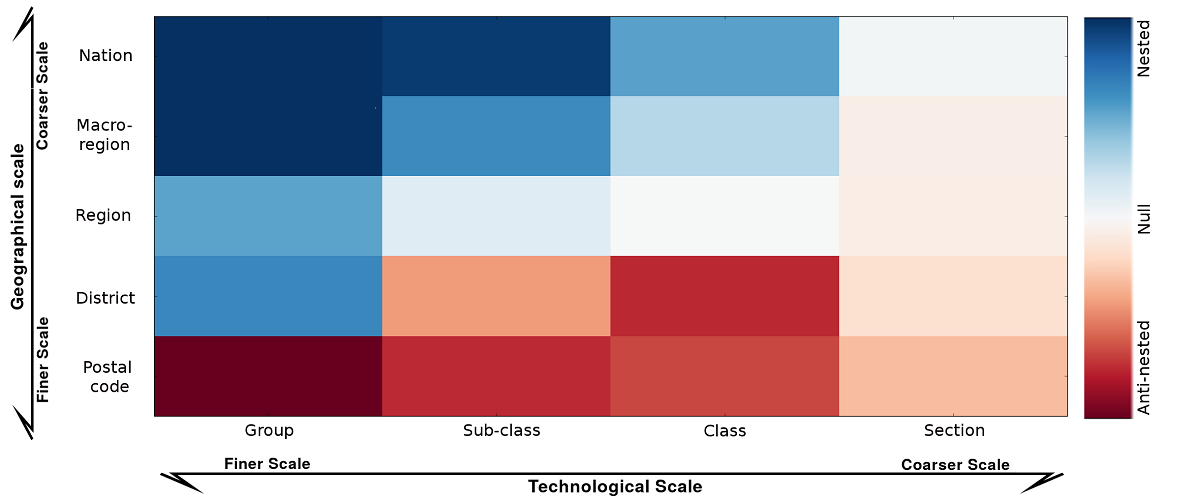}
 \caption{\textbf{Statistical significance of measured Innovation Map temperatures at all scale combinations.} 
 Each row reflects the analysis conducted at a specific geographical aggregation, while each column reflects the analysis conducted at a specific technological aggregation.  The color of the graph spans from blue to red for empirical matrices that show higher (blue) or lower (red) degree of nestedness than expected under the ReCap null model.  The darker the color, the higher the statistical significance with respect to the null model.  For example, the upper left blue cell in the array indicates that the system is significantly more nested (i.e. more diversified) than a random case in which the null matrix has as many rows as the number of nations included in the data and as many columns as the number of technology codes defined according to the finest classification. 
 The above results are robust across time and also hold under different definitions of the patent sample and different binarization strategies for the $IMs$.}
 \label{fig:temps}
\end{figure}

If we perform the same exercise as above but let the technological scale vary while keeping the geographical dimension fixed, we observe a similar yet mirrored situation, in which \textit{zooming-out} produces a structure that is less and less significantly nested and more specialized than what is predicted by the null model. 
For example, at the regional level of geographical aggregation, we observe that nestedness and hence diversification are higher at a finer technological scale, while the system is significantly less nested and more specialized at the broadest level of technological aggregation. 
The overall picture that we obtain is that diversification prevails in the upper-left corner of the matrix, while specialization dominates in the lower-right corner. 
In other words, the scales at which the phenomenon is observed define a \textit{frontier} separating the scale combinations that are significantly more nested than would be expected given the nestedness of a more coarse-grained representation of the same system from the scale combinations that are significantly less nested.
Figure \ref{fig:temps} summarizes our findings.

\section{Discussion}

It is well known in economics that diversification is typically observed in macroeconomic aggregates, while specialization characterizes the behavior of microeconomic agents. 
However, while intuition suggests that countries are highly diversified entities and that the firms constituting their productive tissue are bound to a higher degree of specialization due to capacity constraints, it is not a trivial task to locate the boundary separating diversified scales characterized by complex interactions between capabilities from specialized scales at which capabilities are clustered in relatively independent sets. Our work adds insight in this direction by showing that there are significant interactions between geographical and technological scales and we demonstrate that the prevalence of specialization \emph{vis \`{a} vis} diversification in the fittest locations depends on the scale of observation. 

In particular, we can measure whether the aggregated capabilities linking locations to technologies at a given scale are nested -- suggesting a complex (evolutionary) behavior of the system -- or rather the result of essentially independent activities allowing to reduce the system \emph{at that scale} to a set of disjoint non interacting sub-systems. For example, we find that at the regional level of aggregation, locations are specialists if we observe the system at a coarser technological scale (e.g. IPC Section) but they become generalists if we observe the system at a finer technological scale (e.g. IPC groups). Therefore, diversification and specialization are not elements of a dichotomy, but rather the extremes of a continuum that is uncovered when the geographical and technological granularity of the 
innovation system 
are defined appropriately.

Our study has direct implications for the literature on National and Regional Systems of Innovation \cite{freeman1989technology,cooke2001regional} and can be used to inform policy makers about the dependence of the technological scope to be pursued depending on the geographical scale at which they operate. 
More in general, although the present analysis focuses on technological innovation, our methodology can be generalized to a broader class of social and natural systems characterized by emergent interactions between aggregates at different scales (e.g. geographical, technological, temporal, etc..) to uncover the seeds of irreducible complexity.


\newpage

\bibliographystyle{unsrt}
\bibliography{bibliography}

\newpage

\end{document}